\documentclass[twocolumn,prb,aps,amsfonts]{revtex4-1}
\usepackage{latexsym}
\usepackage{dcolumn}
\usepackage[dvips]{graphicx}
\usepackage{amssymb}
\usepackage{graphicx}
\usepackage{psfrag, color}
\usepackage{verbatim}
\usepackage{amsmath}
\usepackage{epsf}

\newcommand{\vk}{{\bf k}}
\newcommand{\vek}{{\varepsilon_k}}

\newcommand{\vekp}{{\varepsilon_{k'}}}

\begin{document}

\draft

\title{Screening and transport in 2D semiconductor systems at low temperatures} 
\author{S. Das Sarma$^1$ and E. H. Hwang$^{2}$}
\address{$^1$Condensed Matter Theory Center, 
Department of Physics, University of Maryland, College Park,
Maryland  20742-4111 \\
$^2$SKKU Advanced Institute of Nanotechnology and Department of Physics, Sungkyunkwan
University, Suwon, 440-746, Korea} 
\date{\today}

\begin{abstract}
Low temperature carrier transport properties in two-dimensional (2D) semiconductor systems can be theoretically well-understood 
within a mean-field type RPA-Boltzmann theory
as being limited by scattering from screened Coulomb disorder arising from random quenched charged impurities in the environment. In the current work, we derive a number of simple analytical formula, supported by realistic numerical calculations, for the relevant density, mobility, and temperature range where 2D transport should manifest strong 
intrinsic (i.e., arising purely from electronic effects and not from phonon scattering)
metallic temperature dependence in different semiconductor materials arising entirely from the 2D screening properties, thus providing an explanation for why the strong temperature dependence of the 2D resistivity can only be observed in high-quality 
and low-disorder (i.e., high-mobility) 2D samples 
and also why some high-quality 2D materials (i.e., n-GaAs) manifest much weaker metallicity than other materials. We also discuss effects of interaction and disorder on the 2D screening properties in this context as well as compare 2D and 3D screening functions to comment why
such a strong intrinsic temperature dependence arising from screening cannot occur in 3D metallic carrier transport.
Experimentally verifiable predictions are made about the quantitative magnitude of the maximum possible low-temperature metallicity in 2D systems and the scaling behavior of the temperature scale controlling the quantum to classical crossover where the system reverses the sign of the temperature derivative of the 2D resistivity at high temperatures.

\end{abstract}

\maketitle

\section{introduction}

The observation of a strong apparent metallic temperature dependence of the 2D electrical resistivity in high-quality (i.e., low-disorder) semiconductor systems at low carrier densities has become fairly routine \cite{one,one1} during the last 20 years ever since the first experimental report of such an effective metallic behavior in high-mobility n-Si MOSFETs \cite{two}.
Typically, the 2D resistivity $\rho(n,T)$, where $n$ is the 2D carrier density and $T$ is the temperature, increases with increasing temperature by a substantial fraction in the 0.1K -- 5K regime at ``intermediate" carrier densities before phonon effects become operational at higher temperatures. At very low density, the system becomes a disorder-driven strongly localized  
insulator with an activated (or variable-range hopping) resistivity whereas at high density, the metallic temperature dependence is suppressed with the resistivity being essentially temperature-independent (except perhaps for weak localization effects at very low temperature \cite{three} which we ignore in the current work). 
The 2D metallic temperature dependence being of interest here arises from intrinsic electronic effects unrelated to phonon scattering (which produces well-known and well-understood temperature dependence in the carrier resistivity of metals and semiconductors), and thus  the low temperature transport being discussed in the current work refers to the so-called Bloch-Gr\"{u}neisen regime where phonon scattering is strongly suppressed.

The low-density (``insulating") and the high (or intermediate) density (``metallic") transport regimes are separated by a crossover density scale $n_c$ (sometimes refereed to as a critical density although it is really a crossover density scale separating an effective metallic phase for $n>n_c$ from a strongly localized insulating phase for $n<n_c$) which depends on the sample ``quality", decreasing (increasing) with decreasing (increasing) amount of quenched disorder in the system. 
This low-temperature density-driven crossover behavior across $n_c$ in going from an effective strongly insulating phase ($n<n_c$) to an effective metallic phase ($n>n_c$), which is sometimes quite sharp, is often referred to \cite{one} as the two-dimensional metal-insulator-transition (2D MIT) -- a terminology we will use in the current work also although in our picture this is not a quantum phase transition at all, but is simply a sharp crossover from a strongly-localized insulating phase to a weakly-localized metallic phase although the weak localization behavior may not manifest itself until the temperature is unrealistically low.\cite{three}
Although a precise experimental characterization of the sample quality (i.e., the amount of quenched disorder) is challenging because of the unknown nature of the impurity distribution \cite{four}, an approximate characterization is provided by the low-temperature sample mobility ($\mu$) at high carrier density (sometimes referred to as the ``maximum mobility") with higher (lower) sample mobility corresponding to lower (higher) critical density. Experiments clearly indicate that the critical density $n_c$ decreases in a particular material system (e.g., Si(100)-MOSFETs) with increasing sample mobility \cite{five,six}, thus providing a larger range of carrier density ($n > n_c$) where the strong metallic temperature dependence manifests itself,  
but this dependence of the metallic transport  behavior on the sample mobility does not directly carry over to a comparison among different materials -- for example, the metallic behavior is strong (weak) for 2D electrons in Si(100)-MOSFETs (n-GaAs) for $\mu \sim 2\times 10^4$ ($2\times 10^6$) cm$^2$/Vs. Thus, the necessary high mobility for the manifestation of strong metallic temperature dependence in the 2D transport properties depends strongly on the materials system under consideration although in a given 2D system [e.g., Si (100) MOSFETs], 
the metallicity is typically enhanced with increasing mobility. Clearly, having a high mobility (low disorder) is a necessary, but not a sufficient, condition 
for the manifestation of a strong metallic temperature 
dependence in the 2D resistivity. Similar to the mobility, the density and the temperature range for the manifestation of the 2D metallic transport is nonuniversal and strongly materials dependent although within the same material system, the temperature dependence is stronger (weaker) with decreasing (increasing) density as long as $n>n_c$ is satisfied. For example, in n-Si(100) MOSFET (n-GaAs), metallicity is observed for $n\sim 10^{11}$ ($10^9$) cm$^{-2}$ 
in spite of the mobility of the GaAs system being typically two orders of magnitude higher!

The current work is focused on analytical understanding of the various materials parameters which are necessary (and sufficient) for the manifestation
of the strong 2D metallic behavior as reflected in the temperature dependent resistivity of 2D semiconductor carriers.
The theory developed in this article is based on the highly successful mean field model of the metallic temperature dependence in the 2D resistivity as arising from the screened Coulomb disorder in the semiconductor through the strong temperature dependence of 2D screening.
The problem is complex 
even at the mean field level where electron-electron interaction is treated entirely through static RPA screening of disorder
because the total number of independent physical parameters is large. In addition to carrier density ($n$), temperature ($T$), and mobility ($\mu$) mentioned above, transport in 2D systems depends also on carrier effective mass ($m$), background lattice dielectric constant ($\kappa$), valley ($g_v$) and spin ($g_s$) degeneracy of the 2D materials, various materials parameters characterizing electron-acoustic phonon scattering in the system (phonon velocity, Bloch-Gr\"{u}neisen temperature, deformation potential coupling, piezoelectric coupling, etc.) determining the phonon scattering contribution to the electrical resistivity (which is, by definition, temperature dependent and must be negligible in order for the screening induced temperature dependence to be observable), and finally the detailed impurity distribution characterizing the system disorder (with the maximum mobility being the minimal parameter defining the system disorder). Given this large a set of relevant independent parameters affecting 2D transport properties, it seems at first hopeless that anything sensible can be stated analytically about the necessary and sufficient conditions for the manifestation of 2D metallicity. We show in the current work, however, that a few effective parameters actually define the theory reasonably well, providing an excellent qualitative picture for when and where one expects the 2D resistivity to manifest a strong metallic temperature dependence. We also present detailed numerical transport results for $\rho(T,n)$ in several representative 2D systems within the RPA-Boltzmann mean field theory in support of our qualitative analytical results.

The rest of the paper is organized as follows. In section II we provide a brief comparative discussion of 2D and 3D temperature dependent screening properties  of electron liquids within RPA to emphasize the physical origin of the strong temperature dependence of 2D resistivity as limited by scattering from screened Coulomb disorder. In section III we present our main analytical arguments deriving the conditions for strong 2D metallicity and emphasizing the key role of the dimensionless parameters $q_{TF}/2k_F$ and $T/T_F$, where $q_{TF}$, $k_F$, $T_F$ are respectively the 2D Thomas-Fermi screening constant, the 2D Fermi wave number ($k_F \propto \sqrt{n}$), and the Fermi temperature ($T_F \propto n$) in determining 2D metallicity. We also provide direct numerical results for $\rho(T,n)$ to support our analytical results in Section III. In section IV we theoretically consider possible corrections to the 2D screening function arising from disorder and electron-electron interaction effects. We conclude in section V with a summary of our results, and discussing open questions and possible future directions.

\section{Transport and Screening}

The density and temperature dependent 2D conductivity limited by disorder scattering is given within the Boltzmann transport theory by
\begin{equation}
\sigma = \frac{ne^2 \langle \tau \rangle}{m},
\label{eq1}
\end{equation}
where the transport relaxation time, $\langle \tau \rangle = \langle \tau(T,n) \rangle$, is defined by the thermal averaging
\begin{equation}
\langle \tau \rangle =  {\sum_{\vk} \vek \tau(k) \left (-\frac{\partial f(\vek)}{\partial \vek} \right ) }  \bigg /
{\sum_{\vk} \vek \left (-\frac{\partial f(\vek)}{\partial \vek} \right ) } ,
\label{eq2}
\end{equation}
with $\vek = \hbar^2 k^2/2m$ the noninteraction kinetic energy, $k=|\vk|$ the 2D wave number, and $f(\vek)$ is the Fermi distribution function. In Eq.~(\ref{eq2}), an integral over the 2D wave vector $\vk$ is implied by the summation. The wave vector dependent relaxation time $\tau(k)$ is given by the Born approximation treatment of disorder scattering \cite{one1,andormp}
\begin{equation}
\frac{1}{\tau(k)} = \frac{2\pi n_i}{\hbar} \sum_{\bf k'}  \left |u({\bf k-k'}) \right
  |^2  (1-\cos\theta) \delta(\vek-\vekp),
\label{eq3}
\end{equation}
where $\vk$, $\vk'$ are the incident and the scattered carrier wave vectors (and $\theta$ the angle between them) with the $\delta$-function ensuring energy conservation due to elastic scattering by random quenched charged impurities with an effective 2D concentration of $n_i$ per unit area. The carrier-impurity scattering potential is given by the screened Coulomb disorder $u({\bf q})$ defined as
\begin{equation}
u({\bf q}) = \frac{v({ q})}{\epsilon(q)}, 
\label{eq4}
\end{equation}
where $v(q) = 2\pi e^2/(\kappa q)$ is the 2D Coulomb interaction (with $\kappa$ the effective background lattice dielectric constant) and $\epsilon(q)$, the carrier dielectric screening function, is given within RPA by
\begin{equation}
\epsilon(q) = 1+ v(q) \Pi(q),
\end{equation}
where $\Pi(q)$ is the finite temperature non-interaction 2D polarizability function defined by \cite{andormp}
\begin{equation}
\Pi(q) = g_sg_v \sum_{\vk} \frac{f(\vek) - f(\varepsilon_{|{\bf k+q}|})}{\vek - \varepsilon_{|{\bf k+q}|}}.
\label{eq6}
\end{equation}
We will not discuss much the theoretical details for the RPA-Boltzmann transport theory for disorder scattering as provided in Eqs.~(\ref{eq1})--(\ref{eq6}) above since it has already been extensively discussed by us in the literature \cite{one1}. We note that the actual quantitative theory takes into account the realistic quasi-2D nature of the semiconductor system by incorporating appropriate form factors in the Coulomb interaction and the Coulomb disorder using the realistic quasi-2D confinement wavefunctions of the 2D carriers. Also, for 3D systems, Eqs.~(\ref{eq1})--(\ref{eq6}) apply equally well except, of course, for the wave vector $\vk$ being 3D and integrals in Eqs.~(\ref{eq2}), (\ref{eq3}), and (\ref{eq6}) being three-dimensional with the 3D Coulomb interaction being given by $4\pi e^2 /(\kappa q^2)$.

\begin{figure*}[t]
	\centering
	\includegraphics[width=1.8\columnwidth]{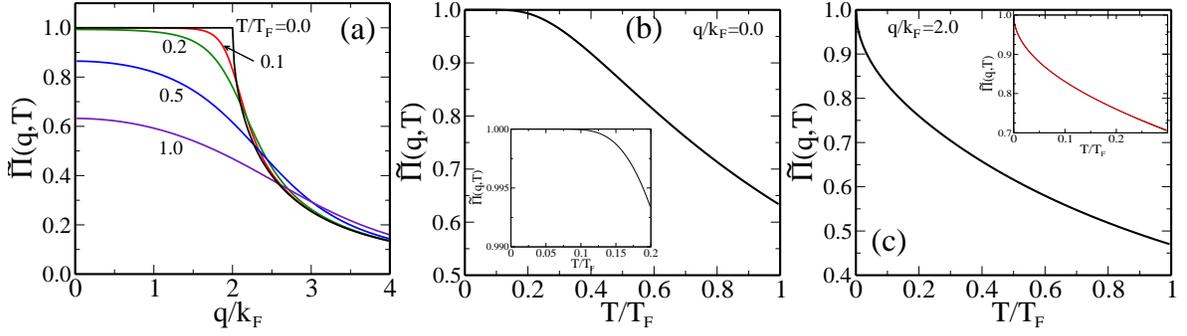}
	\caption{
(a) 2D polarizability $\tilde{\Pi}(q,T)=\Pi(q,T)/N_F^{2D}$ as a function of wave vector for various temperatures, $T=0$, 0.1, 0.2, 0.5, 1.0$T_F$. (b) 2D polarizability as a function of temperature at $q=0$. Inset shows $\tilde{\Pi}(q=0,T)$ at low temperatures. The asymptotic form for $T/T_F \ll 1$ is given by $\tilde{\Pi}(q=0,T) = [1-\exp[-T_F/T]$.
(c) 2D polarizability as a function of temperature at $q=2k_F$. Inset shows $\tilde{\Pi}(q=2k_F,T)$ at low temperatures. The asymptotic form for $T/T_F \ll 1$ is given by $\tilde{\Pi}(q=2k_F,T) = 1-\sqrt{\frac{\pi}{4}}(1-\sqrt{2})\zeta(1/2) \sqrt{T/T_F}$, where $\zeta(x)$ is the Riemann zeta function.
}
\label{fig1}
\end{figure*}

\begin{figure*}[t]
	\centering
	\includegraphics[width=1.8\columnwidth]{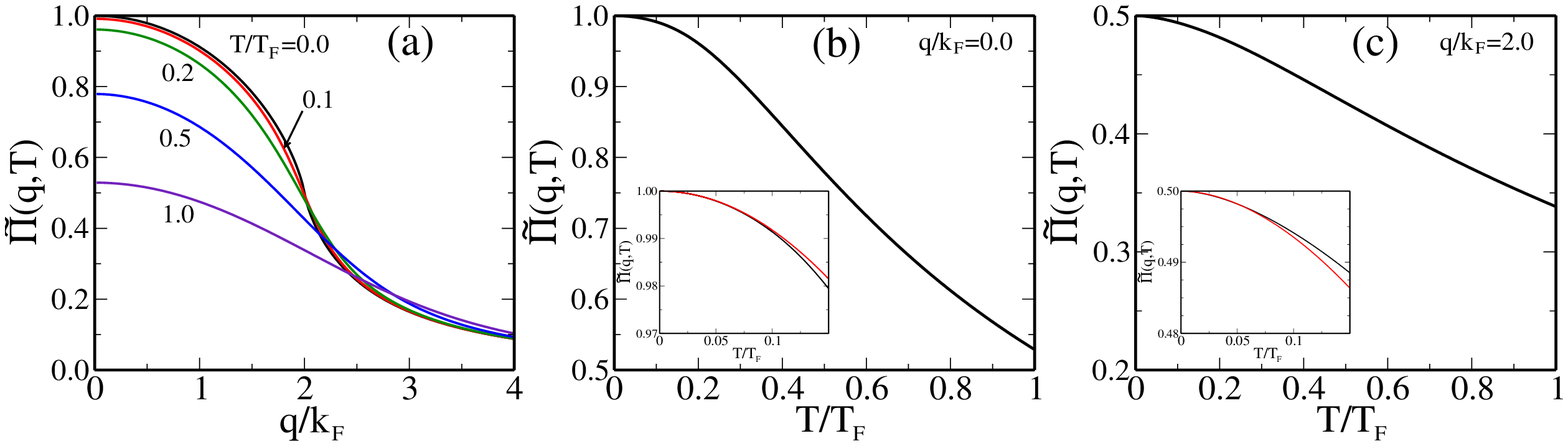}
	\caption{
(a) 3D polarizability $\tilde{\Pi}(q,T)=\Pi(q,T)/N_F^{3D}$ as a function of wave vector for various temperatures, $T=0$, 0.1, 0.2, 0.5, 1.0$T_F$. (b) 3D polarizability as a function of temperature at $q=0$. Inset shows $\Tilde{\Pi}(q=0,T)$ at low temperatures. The asymptotic form for $T/T_F \ll 1$ is given by $\tilde{\Pi}(q=0,T) = 1- \frac{\pi^2}{12} (T/T_F)^2$ (red line).
(c) 3D polarizability as a function of temperature at $q=2k_F$. Inset shows $\tilde{\Pi}(q=2k_F,T)$ at low temperatures. The asymptotic form for $T/T_F \ll 1$ is given by $\tilde{\Pi}(q=2k_F,T) = \frac{1}{2} - \frac{\pi^2}{48}(\frac{T}{T_F})^2[1-\ln\frac{T}{T_F}]$ (red line).
}
\label{fig2}
\end{figure*}

To understand the role of screening in determining 2D transport behavior, it is important to realize that the most resistive carrier scattering is the $2k_F$ back-scattering (i.e., $|\vk-\vk'| = 2k_F$) where an electron on the Fermi surface gets scattered backward (with a scattering angle $\theta = \pi$) by disorder. Thus, the dominant contribution to 
the temperature dependence of the resistivity at low temperatures comes from the behavior of the screening  function $\Pi(q)$ around $q \approx 2k_F$. Sometimes the $2k_F$ scattering is referred to as the scattering from Friedel oscillations \cite{twelve} because the singularity structure of the polarizability function at $q=2 k_F$ (the so-called Kohn anomaly \cite{thirteen}) translates to real space Friedel oscillations \cite{twelve} of the screened potential.

This is true both in 2D and 3D, and we, therefore, show in Figs.~1 and 2 respectively the calculated temperature dependent screening 
(or equivalently, the noninteracting polarizability)
function in 2D and 3D in dimensionless units [$\tilde{\Pi}(q,T) = \Pi(q,T)/\Pi(0,0)$].   A comparison of the two figures (Figs.~1 and 2) clearly brings out the key importance of $2k_F$ screening in determining the 2D metallic temperature dependence in the disorder-limited carrier resistivity,
as was already pointed out by Stern quite a while ago \cite{sternprl1980}.
The temperature dependence of the Friedel oscillations in the screening clouds around the charged impurity centers turns out to be very strong (weak) in 2D (3D) electron systems as shown in Figs.~\ref{fig1} and \ref{fig2} here and discussed below.

First, we note that the 2D screening function (Fig.~1) is very strongly (going as $\sqrt{T/T_F}$) thermally suppressed at $q \approx 2k_F$ compared with very weak (going as $e^{-T_F/T}$) suppression at long wavelength ($q=0$). This low-temperature thermal suppression of $q \approx 2k_F$ screening in 2D systems is the underlying physical mechanism leading to the strong metallic resistivity in 2D systems \cite{sternprl1980,sternsse1985,goldprb1986,zalaprb2001,sarmaprb2004}. We note that the often used long-wavelength screening approximation (i.e. the Thomas-Fermi approximation), although well-valid at $T=0$ since 
the 2D screening function is constant at $T=0$ for $0 \leq q \leq 2k_F$ by virtue of the constant energy independent 2D density of states, fails completely for the calculation of 2D resistivity at finite temperatures since it predicts a very weak temperature-dependent 2D resistivity for $T \ll T_F$ whereas the full wave vector dependent polarizability, which includes the anomalous $\sqrt{T/T_F}$ suppression of screening around $q \approx 2k_F$, predicts a strong linear-in-$T/T_F$ increase of the metallic 2D resistivity at low temperatures. \cite{sternprl1980,sternsse1985,goldprb1986,zalaprb2001,sarmaprb2004}
This strong temperature-dependence of the 2D $2k_F$ screening function is the mechanism 
underlying strong metallicity in 2D semiconductor systems at intermediate densities where the value of $T/T_F$ is not necessarily small leading therefore to a substantial screening dependent thermal effect.  Physically, with increasing temperature, the screened Coulomb disorder, particularly for the important scattering wavenumbers around $2k_F$, is being enhanced strongly due to thermally suppressed screening, leading to an enhanced resistivity due to impurity scattering.

Second, the 3D screening function in Fig.~2 has qualitatively different temperature dependence 
compared with the 2D screening function in Fig.~1. In fact, the temperature dependence of the 3D screening function obeys the ``expected" Sommerfeld expansion behavior in the sense that the low-temperature suppression of screening is a weak quadratic correction going as $O(T/T_F)^2$. This weak quadratic temperature dependence applies both for long-wavelength Thomas-Fermi screening ($q=0$) as well as for $2k_F$-screening ($q=2k_F$) implying weak temperature dependence introduced in the 3D resistivity for $T/T_F \ll 1$ in sharp contrast to the 2D system where the anomalous $O(\sqrt{T/T_F})$ temperature dependence of screening at $q=2k_F$, which violates the Sommerfeld expansion, leads to a strong temperature dependence in the carrier resistivity.
Thus, the key to understanding the strong metallic temperature dependence in the 2D resistivity is the non-analytic temperature dependence of the 2D polarizability arising from the cusp at $2k_F$ in the non-interacting 2D polarizability leading to the failure of the Sommerfeld expansion. 
\cite{sternprl1980,sternsse1985,goldprb1986,zalaprb2001,sarmaprb2004}

For the sake of completeness we quote below the leading order analytical temperature-dependence of the polarizability function in 2D and 3D systems, whereas in Figs.~1 and 2 the full numerically calculated polarizability is shown for arbitrary temperatures:
\begin{equation}
\tilde{\Pi}_{2D}(q=2k_F,T) = 1-\sqrt{\frac{\pi}{4}}(1-\sqrt{2})\zeta(\frac{1}{2})\sqrt{\frac{T}{T_F}}, 
\label{eq7}
\end{equation}
where $\zeta(x)$ is the Riemann zeta function.
\begin{equation}
\tilde{\Pi}_{2D}(q=0,T) = 1- \exp(-T_F/T).
\label{eq8}
\end{equation}
\begin{equation}
\tilde{\Pi}_{3D}(q=2k_F,T) = \frac{1}{2} - \frac{\pi^2}{48}\left ( \frac{T}{T_F} \right )^2 \left (1- \log \frac{T}{T_F} \right ).
\label{eq9}
\end{equation}
\begin{equation}
\tilde{\Pi}_{3D}(q=0,T) = 1 - \frac{\pi^2}{12} \left ( \frac{T}{T_F} \right )^2.
\label{eq10}
\end{equation}
In Eqs.~(\ref{eq7})--(\ref{eq10}), $T_F = E_F/k_B$ is the Fermi temperature,  and $\tilde{\Pi}_{2D} = \Pi(q,T)/N_F^{2D}$ and $\tilde{\Pi}_{3D} = \Pi(q,T)/N_F^{3D}$, where $N_F^{2D} = \Pi_{2D}(q=0,T=0)$ and $N_F^{3D} = \Pi_{3D}(q=0,T=0)$ are the 2D and 3D density of states, respectively.

Before concluding this section, we emphasize that screening is a vital mechanism for 2D semiconductor transport because the disorder in the semiconductor environment arises primarily from random quenched charged impurities whose long-range Coulomb potential must be screened for reasonable theoretical results.  Thus, within a physical mean field approximation, the 2D charge carriers (electrons or holes) are scattered from the screened Coulomb disorder, and therefore, any strong temperature dependence in the screening function, particularly for $2k_F$-scattering which dominates transport at lower temperatures, must necessarily be reflected in the 2D resistivity.

\section{Theory and Numerical Results}

Having established the importance of 2D screening in producing the strong metallic temperature dependence, we now analytically derive a number of conditions constraining the magnitude of the metallic temperature dependence of 2D transport properties which would explain the materials dependence of the metallic behavior as well as provide reasons for why this metallic behavior remained essentially undiscovered (although there were occasional hints \cite{champrl1980}) until the 1990s in spite of there being numerous experimental investigations of 2D semiconductor transport properties in the 1970s and 1980s. \cite{andormp}

\begin{figure}[t]
	\centering
	\includegraphics[width=.8\columnwidth]{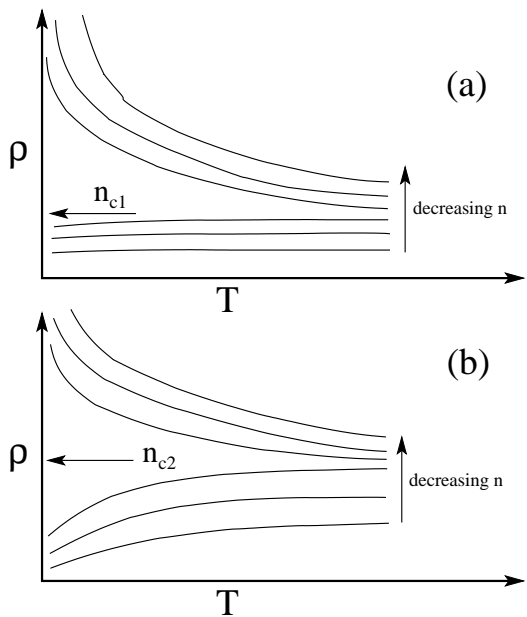}
	\caption{Schematic $\rho(T)$ behavior (for various values of 2D carrier density $n$) for low-mobility (a) and high-mobility (b) systems.  The figure shows the high $n_c$ (a) and low $n_c$ (b) (i.e., $n_{c1} > n_{c2}$) with weak (strong) temperature dependence in $\rho(T)$ in the metallic phase ($n>n_c$) in (a) [(b)] and with very similar exponential insulating temperature dependence in the localized phase ($n<n_c$).
\label{fig0}
}
\end{figure}

In Fig.~\ref{fig0} we schematically depict the two distinct generic
experimentally-observed situations for 2D 
$\rho(T,n)$ with
Fig.~\ref{fig0}(a) and (b) respectively showing the resistivity
$\rho(T)$ for various density ($n$) in low-mobility (high-disorder)
and high-mobility (low-disorder) situations. The only difference
between the two situations is that one [Fig.~\ref{fig0}(a)] has a
``high" value of $n_c$ (because of stronger disorder) whereas the
other [Fig.~\ref{fig0}(b)] has a ``low" value of $n_c$ (because of
weaker disorder). Thus, Figs.~\ref{fig0}(a) and (b) qualitatively show
the respective 2D MIT behaviors in the early ($<1995$) \cite{andormp} and the present
($>1995$) \cite{one} days
or in low-mobility 2D systems \cite{new} and in high-mobility systems \cite{new2}, respectively.
In Fig.~\ref{fig0}(a) and (b) the temperature dependence of $\rho(n,T)$
is weak and strong respectively for $n>n_c$. We mention in this
context the seminal importance of the work of Kravchenko and
collaborators \cite{two,kravchenko1995} who first experimentally
established the connection between the sample quality and the strong
temperature dependence of the 2D resistivity in the metallic ($n >
n_c$) phase using low-temperature transport studies in high-mobility
($>10,000$ cm$^2$/Vs) Si-MOSFETs. Indeed, it is the seminal $1994-95$
work of Kravchenko and collaborators which created the modern subject
of 2D MIT, serving as the temporal milestone separating the early days
of 2D MIT \cite{andormp} [i.e., Fig.~\ref{fig0}(a)] from the present days [i.e.,
  Fig.~\ref{fig0}(b)] of 2D MIT \cite{one}.  
We emphasize that both Figs.~3(a) and (b) manifest essentially identical strongly localized 
insulating phase for $n<n_c$, but differ in the temperature dependence of the effective metallic phase with (older) lower mobility samples [Fig.~ 3(a)] showing little temperature dependence for $n>n_c$ and the (newer) higher mobility samples manifesting strong metallic temperature dependence [Fig.~3(b)] for $n>n_c$.   
Below we establish that the key to the strong metallic temperature
dependence of the 2D resistivity (for $n>n_c$) is having
(low-disorder-induced) low values of the crossover density $n_c$,
which makes $\rho(T)$ manifest somewhat complementary temperature
dependence ($d\rho/dT >0$ for $n \agt n_c$ and $d\rho/dT <0$ for $n <
n_c$) on two sides of the 2D MIT
as depicted in Fig.~3(b).  On the other hand, for low-mobility samples where $n_c$ is necessarily high, the metallic phase (for $n>n_c$) does not manifest any intrinsic temperature dependence [as shown in Fig.~3(a)] except at high enough temperatures where phonon scattering effects 
(ignored in the current work)
become important.
We emphasize, however, that at very
high (low) density both kinds of samples (low and high disorder in
Fig.~\ref{fig0}) manifest similarly weak (strong) temperature
dependence. 
We focus only on the metallic ($n>n_c$) phase using the RPA-Boltzmann theory and discuss the necessary and sufficient conditions for the manifestation of a strong 
intrinsic (i.e. not phonon-related)
temperature dependence in the 2D resistivity. The transition to the insulating phase has been discussed by us elsewhere \cite{six,tracyprb2009}
and is not a part of the current work where the focus is entirely on the effective metallic regime of $n>n_c$.

To understand how the strong (weak) metallic temperature dependence
(for $n>n_c$) correlates with low (high) values of $n_c$, we introduce
three independent density dependent temperature scales ($T_F$,
$T_{BG}$, $T_D$) which characterize the temperature dependence of the
resistivity in the metallic phase. These are the electron temperature
scale defined by the Fermi temperature ($T_F$), the phonon temperature
scale defined by the  Bloch-Gr\"{u}neisen temperature ($T_{BG}$), and
the disorder temperature scale defined by the Dingle temperature: 
\begin{equation} 
k_BT_F = E_F = \frac{\hbar^2 k_F^2}{2m} = \frac{\hbar^2}{2m}\left ( \frac{4\pi n}{g_s g_v} \right ) \propto n,
\end{equation}
\begin{equation}
k_B T_{BG} = 2 \hbar k_F v_{ph} = 2 \hbar v_{ph} \left ( \frac{4\pi n}{g_s g_v} \right )^{1/2} \propto n^{1/2},
\end{equation}
\begin{equation}
k_BT_D = \Gamma = \frac{\hbar}{2} \left (\frac{e}{m\mu} \right ) \propto \mu^{-1}.
\label{dingleT}
\end{equation}
Here $E_F$, $k_F = (4\pi n/g_s g_v)^{1/2}$, $m$, $v_{ph}$, and
$\Gamma$ are respectively the 2D Fermi energy, 2D Fermi wave vector,
the carrier effective mass, the phonon velocity, and the
impurity-scattering induced level broadening (with $\mu$ as the sample
mobility). 
For simplicity, we have defined the level broadening $\Gamma = \hbar/2\tau$ where $\tau$ is the transport relaxation time defining the 2D mobility $\mu = \sigma/ne= e \tau /m$ with $\mu$ being the maximum mobility -- 
in general, the broadening $\Gamma$ (and therefore the Dingle temperature $T_D$) is density-dependent through the density dependence of mobility which is a complication we ignore for our definition of $T_D$.  [We also mention that often the Dingle temperature is defined with an additional factor of $\pi$ in the denominator giving a smaller value for $T_D$ in Eq. (\ref{dingleT}).]
To keep our considerations general, we assume a carrier
valley degeneracy $g_v$ and a spin degeneracy $g_s$ so that the total
ground state degeneracy is $g_sg_v$ --- $g_s=2$ in general except in
the presence of a strong applied magnetic field which could
spin-polarize the system making $g_s=1$ whereas $g_v=1$ in general
except in Si-MOSFETs where other values of $g_v>1$ are possible because of the peculiar 
multi-valley Si
bulk conduction band structure.  
The Fermi temperature $T_F$ defines the intrinsic quantum temperature
scale for the 2D electrons, and when $T_F$ is very large (i.e., $n$
very high since $T_F \propto n$), there cannot be any temperature
dependence in the metallic resistivity at low temperatures arising
from intrinsic electronic effects since $T/T_F \ll 1$. Thus, $n_c$
needs to be relatively low just in order to keep $T_F$ low so that
$T/T_F$ is not too small for $n>n_c$ before phonon effects become significant. The
Bloch-Gr\"{u}neisen temperature $T_{BG}$ ($\propto k_F \propto
\sqrt{n}$) defines the characteristic temperature scale ($T>T_{BG}$) for phonon
scattering effects to become important in the 2D metallic
resistivity. For $T<T_{BG}$, phonon effects are strongly suppressed,
leading to a weak $T^p$-type ($p \approx 5-7$) very high power law in
the 2D resistivity arising from phonon scattering whereas for
$T>T_{BG}$, the phonon scattering contribution to the 2D resistivity
is linear in $T$ (which is universally observed in all 2D
semiconductor systems in the metallic phase for $T>1-10K$ depending on
the carrier density).  
Thus, the observation of 2D metallic behavior at low temperatures requires $T<T_{BG}$ since trivial phonon scattering contribution to the resistivity 
for $T>T_{BG}$, which is always present,
is not the issue here.

This immediately implies that $T_F <T_{BG}$ is necessary 
for the manifestation of the strong metallic temperature
dependence in the resistivity since otherwise (i.e., for $T_{F} \gg
T_{BG}$) the low-temperature (i.e., $T<T_F$) resistivity will be
already dominated by the $\rho \sim T$ behavior arising from phonon
scattering effects applicable for $T>T_{BG}$. Since $T_F \propto n$
and $T_{BG} \propto \sqrt{n}$, the condition $T_{F} < T_{BG}$
necessitates a low carrier density leading to the conclusion that a
large $n_c$ would lead to the temperature dependence of metallic
resistivity ($n>n_c$) being dominated by phonon scattering
effects. 
This, in fact, typically happened in the 2D systems studied in the 1970s and 1980s 
where phonon effects dominated the metallic resistivity ($n>n_c$) suppressing all intrinsic screening-induced temperature effects \cite{andormp}.
Thus, simple dimensional considerations of the characteristic
electronic ($T_F$) and phononic ($T_{BG}$) temperature scales in the
problem lead to the inevitable conclusion that any strong metallic
temperature dependence arising purely from a quantum electronic
mechanism necessitates $T_F < T_{BG}$ (or at least $T_F \gg T_{BG}$ is
not allowed), and hence necessarily a low $n_c$ (so that $T_F$ is not
too large even for $n \agt n_c$). As an aside we mention that in 3D
metals $T_F \sim 10^4 K$ and the phonon temperature scale $T_{BG}$ is
replaced by the Debye temperature $\Theta_D \sim 10^2 K$, so that $T_F
\gg \Theta_D$ always. This means that the metallic temperature
dependence in the resistivity arising purely from an electronic
mechanism is simply impossible in 3D metals at low temperatures where
phonon effects always dominate down to low temperatures. There can be
a weak $T^2$ contribution to the resistivity in 3D metals arising from
electron-electron scattering through umklapp processes which cannot
happen in the 2D semiconductor systems since the umklapp scattering
involves very large lattice-scale momentum transfer not of interest in
semiconductor transport.

The role of the disorder-dependent Dingle temperature $T_D$ in the transport problem is rather subtle and is relevant at the lowest temperatures $T<T_D$ where $T_D$ acts as a lower cutoff suppressing the temperature dependence for $T<T_D$. 
This is because the strong temperature dependence of carrier screening leading to the metallic temperature dependence is cutoff for $T<T_D$ by impurity disorder effects parametrized by the Dingle temperature. 
This is because the strong temperature dependence of the 2D polarizability around $2k_F$ is suppressed for $T \agt T_D$ as disorder rounds off the $2k_F$ screening. \cite{dassarmaprb1986} This is discussed in Sec. IV.
Thus, $T_D$ explains why the metallic temperature dependence for $T<T_F<T_{BG}$ arising from quantum electronic processes does not persist 
(even in the absence of weak localization which is being ignored here)
all the way to $T=0$ as it would for $T_D=0$ (and of course, if the electronic temperature can be reduced indefinitely which may be an impossibility). Thus, the temperature dependence of $\rho(T)$ in the metallic phase ($n>n_c$) is bounded from above (by $T_{BG}$) and from below (by $T_D$) with the screening induced temperature dependence being strong only in the window $T_D \alt T \alt T_F \alt T_{BG}$. Since $T_D \propto \Gamma \propto \mu^{-1}$ where $\mu$ is the characteristic mobility of the system, a large $\mu$ (i.e., low disorder) is necessary in order to keep $T_D$ (as well as $n_c$) small so that the temperature dependence of $\rho(T)$ can show up in an appreciable temperature window satisfying $T_D < T <T_F < T_{BG}$ with $T_{BG} > T_F$ guaranteeing that phonon scattering would not play a role in the 2D MIT physics. Note that if $T_D >T_{BG}$ (i.e., in highly disordered samples) all metallic temperature dependence will be totally suppressed. 
We emphasize that the 2D system must be high quality (i.e., low-disorder and high-mobility) so that both $n_c$ and $T_D$ are small since both $n_c$ and $T_D$ decrease approximately linearly with increasing mobility.
One reason that the metallic temperature dependence in the resistivity manifests itself rather strongly in Si-based 2D systems even for relatively modest values of $\mu$ ($>10,000$ cm$^2$/Vs) is because the effective $T_{BG}$ is rather high in Si because of the high phonon velocity and generally weak electron-phonon coupling.

The high mobility low-disorder samples of Kravchenko {\it et al.} (and of others since then in the modern era of 2D MIT physics) routinely satisfy the constraint $T_D < T < T_F < T_{BG}$ enabling the observation of the strong metallic temperature dependence since $n_c$ and $T_D$ are both low in these high-mobility samples whereas the older Si-MOSFET samples (before the Kravchenko era), where the 2D MIT phenomenon was studied in the early days \cite{andormp,mott1975,adkins1978}, had high disorder (and low mobility) and consequently high $n_c$ (and $T_D$) leading to large $T_F \gg T_{BG}$ (as well as large $T_D >T_{BG}$) in the metallic phase ($n>n_c$) so that no metallicity could be observed except for phonon scattering effects for $T > T_{BG}$. Thus, the amount of disorder in the sample leading to low or high $n_c$ (and $T_D$) is the key to the manifestation of a strong metallic temperature dependence in $\rho(T)$ for $n > n_c$.

The condition derived above, $T_D<T_c<T_{BG}$ where $T_c=T_F$ ($n=n_c$), for the manifestation of the strong metallicity in the 2D system for $n>n_c$ is only a qualitative {\it necessary} condition which allows the intrinsic temperature dependence from the 2D screening effect to show up in transport properties, but whether such a metallic temperature dependence would actually be a quantitatively strong effect or not depends on certain additional {\it sufficient} conditions which we would discuss below.  These sufficient conditions ensure that the $2k_F$-screening is in fact quantitatively significant, not just that it is allowed to be present.

To give a quantitative description underlying the qualitative picture discussed above, we borrow (without any derivations) from our earlier-obtained \cite{one1, sarmaprb2004,sarmaprb2003} theoretical results providing $\rho(T)$ in the 2D effective metallic phase assuming that the resistive scattering arises from screened Coulomb disorder in the system. The quantitative analytical considerations provided below 
for $\rho(T,n)$ in the metallic phase serve three purposes: (1) They reinforce in a concrete manner the qualitative discussion given above establishing how the consideration of the characteristic temperature scales $T_F$, $T_{BG}$, and $T_D$ (particularly, their density and mobility dependence) immediately leads to the conclusion that $n_c$ must be small (i.e., low disorder and high mobility) for the 2D MIT phenomenon to be associated with a strongly metallic temperature dependence in $\rho(T,n)$ for $n > n_c$; (2) they provide a quantitative understanding of what low (or high) $n_c$ actually means in a materials-dependent manner, i.e., tell us how large can $n_c$ be in a specific system (e.g., Si-MOSFETs or 2D GaAs systems) 
and still manifest a strongly temperature-dependent $\rho(T)$ for $n>n_c$ without any phonon effects; and (3) they describe how large or small $n_c$ should be in going from one 2D system to another (e.g., from 2D Si-MOSFETs to 2D GaAs quantum wells) in order for similar metallic temperature dependence to show up in different 2D systems for $n>n_c$.

The Boltzmann transport theory gives \cite{one1,sarmaprb2004,sarmaprb2003} the following analytical results for the semiclassical $\rho_i(T)$ in 2D electron systems at asymptotically low ($T \ll T_F$) and high ($T \gg T_F$) temperatures, respectively
\begin{eqnarray}
\rho_i(T\ll T_F) \approx &\rho_0& \left [ 1   +  \frac{2x}{1+x} \frac{T}{T_F} \right .  \nonumber \\
& + & \left .  y \left ( \frac{T}{T_F} \right )^{3/2} +  O \left ( \frac{T}{T_F} \right )^{2} \right ],
\label{rholow}
\end{eqnarray}
\begin{equation}
\rho_i(T \gg T_F) \approx \rho_1 \frac{T_F}{T}  \left [ 1- \frac{3\sqrt{\pi} x}{4} \left (\frac{T_F}{T} \right )^{3/2} + O\left(\frac{T_F}{T} \right )^3 \right ],
\label{rhohigh}
\end{equation}
where  $x = q_{TF}/2k_F$ and
$y = 2.646  [ x/(1+x)  ]^2$.
In Eqs.~(\ref{rholow}) and (\ref{rhohigh}), $\rho_0 = \rho(T=0)$ and $\rho_1 = (h/e^2) (n_i/n\pi x^2)$
respectively are the impurity-scattering induced semiclassical resistivities (hence $\rho_i$) characterizing the low and the high temperature limits, and $T_F = E_F/k_B$ is the Fermi temperature (with $n$, $n_i$ being the respective 2D carrier density and impurity density in the system, and $k_F=(4\pi n/g_s g_v)^{1/2}$ and $q_{TF} = g_s g_v m e^2/\kappa \hbar^2 $ are the 2D Fermi wave vector and Thomas-Fermi screening wave vector, respectively).
We do not provide the analytical derivations of Eqs.~(\ref{rholow}) and (\ref{rhohigh}), which can be obtained using Eqs.~(\ref{eq1}) -- (\ref{eq10}) as shown in Refs.~\onlinecite{sarmaprb2003,sarmaprb2004}.
While Eqs.~(\ref{rholow}) and (\ref{rhohigh}) provide the metallic contributions to $\rho(T)$ arising from the temperature dependence of the screened Coulomb disorder, the acoustic phonon scattering by itself contributes also to the temperature dependence \cite{andormp,kawamuraprb1992}
given in the high ($T \gg T_{BG}$) and low ($T \ll T_{BG}$) temperature limits by
\begin{equation}
\rho_a(T\gg T_{BG}) \approx \rho_0 + A_{ph} \left ( \frac{T}{T_{BG}} \right ),
\end{equation}
\begin{equation}
\rho_a(T \ll T_{BG}) \approx \rho_0 + B_{ph} \left ( \frac{T}{T_{BG}} \right )^5,
\end{equation}
where $T_{BG}=2\hbar k_F v_{ph}$ is the Bloch-Gl\"{u}neisen temperature with $v_{ph}$ as the relevant phonon velocity -- the constants $A_{ph}$, $B_{ph}$ depend on the elastic properties of the semiconductor \cite{kawamuraprb1992}.

We immediately note that strong metallicity necessitates $T_{BG} > T_F$, which means that we must have $2\hbar k_F v_{ph} > \hbar^2 k_F^2 /2m$, i.e., $k_F < 4 m v_{ph}/\hbar$. Since $k_F \propto \sqrt{n}$, the observation of metallicity is a low-density phenomenon restricted to $n<8 g_v m^2 v_{ph}^2/\pi \hbar^2 \equiv n_p$ where phonon effects are suppressed.
For $n > n_p$, phonon effects become relevant for transport.

In addition, the screening induced metallic temperature dependence [Eqs.~(\ref{rholow}) and (\ref{rhohigh})] can only apply for $n>n_c$ since, for $n<n_c$, strong localization induced insulating behavior will dominate (and the metallic theory does not apply). Thus, the metallic behavior is only allowed in an intermediate density window $n_c < n <n_p$.
It follows right away that if the sample is so dirty that $n_c \agt n_p$, the metallic behavior simply cannot be observed in an experimental sample under any circumstance at any temperature! Thus, a minimal necessary condition for the manifestation of 2D metallic temperature dependence is that
\begin{equation}
n_c < n_p = 8 g_v m^2 v_{ph}^2/\pi \hbar^2,
\label{nc}
\end{equation}
where $n_c$ is the crossover carrier density for the metal-to-insulator transition. Since $n_c$ obviously increases (decreases) with increasing (decreasing) disorder in the system, a minimal condition for the observation of metallicity is that the system must have low disorder or, equivalently, high mobility, at least satisfying Eq.~(\ref{nc}) above.
We also note that Eq.~(\ref{nc}) implies 
[using the Si(100)-MOSFET materials parameters] an $n_c \alt 1.2 \times 10^{11}$ cm$^{-2}$ for Si(100)-MOSFETs consistent with experimental observations in the sense that the modern 2D MIT era started with the Kravchenko-Pudalov seminal 1994-95 2D transport measurements where the critical density is indeed less than $10^{11}$ cm$^{-2}$ whereas the older MOSFETs manifested an insulating phase (with activated conductivity) at a much higher density of $n_c \agt 10^{12}$ cm$^{-2}$, \cite{andormp,mott1975,adkins1978} where according to our analysis, no temperature-dependent metallic conductivity can be observed except for phonon effects for $T>T_{BG}$.

To see the role of high mobility in the modern 2D MIT phenomenon of current interest more clearly we consider the specific criterion of the impurity scattering induced collisional broadening energy scale defined by the level broadening parameter $\Gamma = k_BT_D$ (where $T_D$ is the Dingle temperature). Using the Ioffe-Regel criterion for calculating $n_c$, \cite{six} our condition discussed above, i.e., $n_c \agt n_p$, becomes equivalent to the condition $T_D < T_F < T_{BG}$ for the unambiguous manifestation of the metallic temperature dependence. Using the fact that $\Gamma = \hbar/2\tau$ and $\mu = e\tau/m$, we then get the following necessary condition on mobility for the manifestation of the metallic phase 
\begin{equation}
\mu > \hbar e/(2k_B m T_F).
\label{muc}
\end{equation}
Using a carrier density $n_c \approx 10^{11}$ cm$^{-2}$, we get for the Si(100)-MOSFETs, $\mu > 21,000$ cm$^2$/Vs. For proportionally lower values of $n_c$, the required minimum mobility would be proportionally higher, again reinforcing the fact that high mobility is a necessary prerequisite for the 2D metallic phase (i.e., $n>n_c$) to manifest a strong temperature dependence in the resistivity. It is reassuring to note that indeed all modern Si-MOS samples showing the canonical 2D MIT behavior after the Kravchenko-Pudalov discovery typically have $\mu \agt 20,000$ cm$^{2}$/Vs.

We note that the above constraints on the critical density [Eq.~(\ref{muc})] and the sample mobility ($\mu$) are only the necessary conditions, which may not be sufficient for the actual manifestation of a strong temperature dependent 2D resistivity on the metallic (i.e., $n>n_c$) side.
For example, the actual quantitative screening effect on $\rho(T,n)$ as defined by Eqs.~(\ref{rholow}) and (\ref{rhohigh}), may simply be too small for experimental observation even if the necessary condition of $n_c<n<n_p$ is satisfied.
To discuss this issue of sufficient conditions we go back to Eq.~(\ref{rholow}) and note that for $\rho_i(T)$ to manifest strong temperature dependence, we must have $x \gg 1$ (at least $x>1$) so that $\frac{T_F}{\rho_0}\frac{d\rho}{dT} \approx 2x/(1+x)$ is not too small. This requires $x = q_{TF}/2k_F > 1$, i.e., $q_{TF} > 2k_F$ which translates to 
\begin{equation}
n \alt n_M = 2 g_v^3 m^2 e^4/\kappa^2 \hbar^4 \pi^2,
\label{ncm}
\end{equation}
where $\kappa$ is the background lattice dielectric constant (assuming $g_s=2$). Obviously $n>n_c$ has to be satisfied for the 2D system to be in the metallic phase, and so metallicity requires the additional sufficient condition of
\begin{equation}
n_c < n_M \alt n, 
\label{eq101}
\end{equation}
with $n_M = 2 g_v^3 m^2 e^4/\kappa^2 \hbar^4 \pi^2$.
For Si(100)-MOSFETs with $g_v=2$ we get $n_M \approx 1.2 \times 10^{12}$ cm$^{-2}$, which is much larger than $n_c \approx 10^{11}$ cm$^{-2}$ for the post-1995 era 2D MOSFET samples manifesting metallicity in the $T<T_{BG}$ regime of temperatures.
This large value of $n_M$, however, does explain why low-mobility 2D Si samples do not manifest any metallicity since $n_c$ is large ($>n_M$) in such lower quality samples.

It is gratifying that simple considerations involving $T_F$, $T_{BG}$, $T_D$, and $q_{TF}/2k_F$ immediately lead to the prediction that in Si(100)-MOSFETs there would be an $n_c$ low enough ($n_c \alt 10^{11}$ cm$^{-2}$) for high-mobility ($\mu \agt 20,000$ cm$^{2}$/Vs) samples to show strong metallic $\rho(T)$ behavior for $n \agt n_c$ exactly as observed experimentally in the post-Kravchenko ($> 1995$) samples whereas in older low-mobility samples with $n_c \sim 10^{12}$ cm$^{-2}$, there would be no metallic $\rho(T)$ behavior (except for phonon effects for $T >T_{BG}$) exactly as seen in lower-mobility MOSFET systems \cite{andormp}.

What about other 2D systems such as high-mobility 2D n-GaAs and p-GaAs systems? Below we briefly discuss quantitative implications of Eqs.~(\ref{rholow}) -- (\ref{eq10}) for 2D GaAs systems with respect to the 2D MIT phenomena.

First, 2D n-GaAs has $m=0.07m_e$, $g_v=1$, $\kappa = 13$, and $v_{ph} = 4 \times 10^5$ cm/s in contrast to Si(100)-MOSFETs (considered above in depth) which have $m=0.19m_e$, $g_v=2$, $\kappa = 12$, and $v_{ph} = 9 \times 10^{5}$ cm/s. Applying Eqs.~(\ref{rholow}) -- (\ref{eq10}) to 2D n-GaAs system, we get
\begin{eqnarray}
n_p & \approx & 1.5 \times 10^{10} cm^{-2}; \;\; n_c < 10^{10}cm^{-2}; \nonumber \\
\mu & \approx & 500,000 cm^2/Vs; \;\; n_M \approx 4 \times 10^{10} cm^{-2}.
\end{eqnarray}
This indicates that one would have to go to very low carrier density, way below $10^{10}$ cm$^{-2}$, to see any metallicity in 2D n-GaAs system. Since $T_F(K) \approx 4 \tilde{n}$ in n-GaAs where $\tilde{n}$ is the carrier density measured in $10^{10}$ cm$^{-2}$, the temperature range ($T<T_F$) for any possible metallic behavior would be well below $1K$.
In addition, $q_{TF}/2k_F\approx 0.4/\sqrt{\tilde{n}}$, which means that $q_{TF}/2k_F = 1$ is reached only for $n \approx 1.6 \times 10^9$ cm$^{-2}$, implying that observing strong metallicity (i.e., relatively latge $d\rho/dT$) in 2D n-GaAs would necessitate going to carrier density in the range of $1-2 \times 10^9$ cm$^{-2}$ and $T < 100$ mK, requiring electron mobility of $10^7$ cm$^2$/Vs. Indeed
there is only one experimental report \cite{lillyprl2003} of observing strong metallic behavior in 2D n-GaAs, and it required an ultrahigh mobility of $10^7$ cm$^{-2}$/Vs and a sample of very low carrier density ($\sim 10^9$ cm$^{-2}$) in agreement with our estimates.

It is easy to convince oneself using Eqs.~(\ref{rholow}) -- (\ref{eq10}) and 2D p-GaAs parameters that for GaAs 2D holes, the metallic behavior should be routinely observable in samples with mobilities of $10^5 - 10^6$ cm$^{2}$/Vs at carrier densities around $10^{10}$ cm$^{-2}$. This is indeed the experimental situation.

Thus, we have established in this section why older Si-MOSFETs did not see 2D MIT phenomenology: It is simply because the sample quality was too low and consequently the critical density was too high,
making it impossible for any screening induced temperature effect to manifest itself before the phonon induced temperature effects show up.
It may be worthwhile to obtain some rough comparative quantitative estimates for the metallic temperature dependence in samples with high and low disorder in order to contrast older and newer Si-MOSFET samples. We provide such a quantitative comparison below for two hypothetical Si-MOSFET samples: A (high disorder) and B (low disorder) with high-density mobilities of $5,000$ cm$^2$/Vs (high $n_c$ for sample A) and $50,000$ cm$^2$/Vs (low $n_c$ for sample B) 

Sample A (high disorder) has $n_c = 10^{12}$ cm$^{-2}$, which, using Eq.~(\ref{rholow}) leads to 
\begin{equation}
\left ( \Delta \rho/\rho_0 \right )_A \alt 0.2,
\end{equation}
where $\Delta \rho = \rho(T_{BG}) - \rho(T=T_D)$ is the temperature induced increase in the metallic resistivity (for $n \agt n_c$) arising from the screening effect.

Sample B (low disorder) has $n_c = 10^{11}$ cm$^{-2}$, which, 
using Eq.~(\ref{rholow}), leads to 
\begin{equation}
(\Delta \rho/\rho_0)_B \agt 1.2.
\end{equation}
Thus, sample A (B) would manifest a less than 20 \% (more than 120 \%) increase in the metallic resistivity (for $n \agt n_c$) between $T_D < T < T_{BG}$ arising from screening effects, clearly establishing that having low (high) values of the crossover density $n_c$ is the crucial element of physics determining strong (weak) metallic temperature dependence in the system. Since the temperature range for metallicity ($T \alt T_{BG} \sim \sqrt{n}$) is much smaller for the lower-disorder sample B, as it has a much lower $T_{BG}^{(B)} \sim 14 K$ compared with $T_{BG}^{(A)} \sim 35 K$ in sample A, the actual manifested temperature dependence would look much stronger in sample B, where $\rho(T)$ will increase by a factor of 2 in the $T=0-10K$ regime compared with only a $<10\%$ increase in $\rho(T)$ for sample A in the same temperature ($1-10K$) range. This simple estimate shows why older lower mobility MOSFET samples, extensively studied in the 1970s and 1980s \cite{andormp} with mobilities around $5,000$ cm$^2$/Vs (or less) never manifested any strong metallic behavior because of their relatively high values of $n_c$ whereas the more recently studied MOSFET samples with mobilities above $20,000$ cm$^2$/Vs (and $n_c \sim 10^{11}$ cm$^{-2}$ or less) always manifest strong metallic temperature dependence in its resistivity. The mystery of the so-called strong 2D metallic behavior is thus connected directly to the relative magnitude of $n_c$ as determined by the 2D sample quality.
We do, however, mention that some Si(100) MOSFET samples with relatively higher mobilities manifested observable metallic temperature dependence in the measured resistivity as far back as in the early 1980s,\cite{champrl1980} but this was more of an exception
since Si MOSEFETs with $\mu>10,000$ cm$^2$/Vs were very rare before 1995.

Recently, a spectacularly strong metallic temperature dependent resistivity was observed \cite{huprl2015} in Si(111)-based 2D electrons with an unprecedented high maximum mobility of $\sim 200,000$ cm$^2$/Vs. This ultra-high-mobility 2D Si(111) electron system has a valley degeneracy of 6, and manifested almost an order of magnitude increase in the metallic resistivity (at $n \sim 6\times 10^{11}$ cm$^{-2}$) in the $T=0.3 - 4$ K in contrast to the 2D Si(100) MOSFETs which typically manifest at best a factor of 3 change in the measured resistivity in a similar temperature window \cite{two}. The strong observed metallicity in this high-mobility Si(111) system arises from its high valley degeneracy $g_v=6$, consistent with the bulk 6-valley minima electronic structure of Si conduction band leading to $g_v=2$ (6) in Si(100) [(111)] 2D systems. In the context of this experimental development\cite{huprl2015} it may be worthwhile to compare Si(100) and Si(111) 2D systems with respect to the various parameters ($n_c$, $n_{BG}$, $n_M$, $x=q_{TF}/2k_F$, $T_F$, $T_{BG}$, $T_D$) defining 2D metallic properties.

Using Eqs.~(11)--(24) incorporating the materials parameters ($m$, $\kappa$, $g_v$, etc.) for Si(100) and Si(111) 2D systems we find
$ {n_{BG}^{(111)}}/{n_{BG}^{(100)}} \approx 7.5$,
${n_M^{(111)}}/{n_M^{(100)}} \approx 70$,
${x^{(111)}}/{x^{(100)}} \approx 10$,
${n_c^{(111)}}/{n_c^{(100)}}  \approx {3 \mu^{(100)}}/{\mu^{(111)}}$,
${T_D^{(111)}}/{T_D^{(100)}} \approx {\mu^{(100)}}/{2\mu^{(111)}}$,
and
${T_F^{(111)}}/{T_F^{(100)}} = {1}/{3}$.
The above considerations show that to obtain the same value of $n_c$ in Si(100) and Si(111) systems necessitates that the Si(111) system has a much larger (at least by 3 times) mobility whereas the density range ($n_M$) upto which the metallicity persists is much higher (by a factor of 70!) in Si(111)
compared with the Si(100) system. 
Most importantly, the large valley degeneracy in the Si(111) system implies an effectively large 
(an order of magnitude larger than Si(100) system for the same 2D carrier density)
value of $x=q_{TF}/2k_F$ producing a very large value of $d\rho/dT$ in the metallic phase leading to a much stronger metallic temperature dependence in the resistivity compared with the Si(100) system ($g_v=2$) exactly as observed experimentally.
Interestingly, our numerical comparison of the Si(111) 2D system with the Si(100) 2D system given above suggests, in agreement with the experiment \cite{huprl2015}, the intriguing dichotomy that while the critical density (and thus the density range where strong metallicity is expected) is higher in the former, the actual temperature dependent fractional change in the resistivity is still considerably higher in the Si(111) system even at this higher absolute density compared with the Si(100) system because of the large valley degeneracy operational on the (111) surface!
This establishes that one cannot compare $n_c$ values between two different 2D materials to conclude about their relative strength of metallicity -- although Si (111) has relatively higher $n_c$, it still has stronger metallicity compared with Si (100) system.

\begin{figure}[t]
	\centering
	\includegraphics[width=1.\columnwidth]{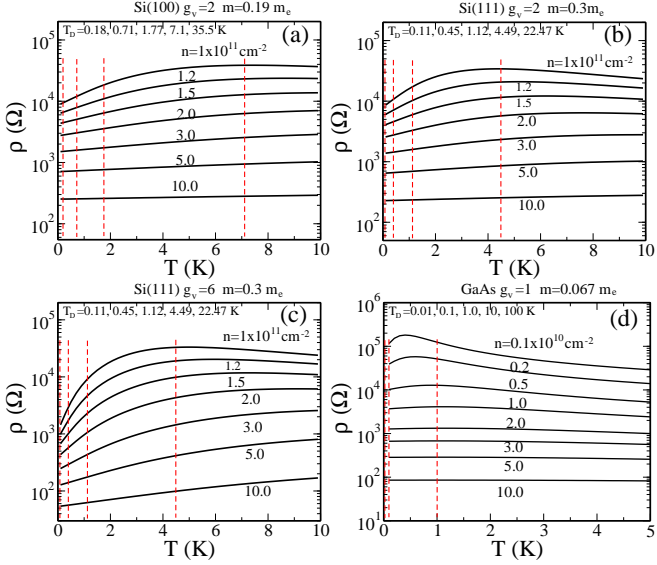}
	\caption{
Resistivity as a function of temperature for different densities (in the unit of $10^{11} cm^{-2}$). The resistivities are calculated with the parameters corresponding to (a) Si(100), (b) Si(111) with $g_v=2$, (c) Si(111) with $g_v=6$, and the same impurity configurations (impurities are located at the interface of Si and SiO$_2$). 
The vertical red dashed lines indicate the Dingle temperatures corresponding to the mobility $\mu = 2\times 10^5$, $5 \times 10^4$, $2 \times 10^4$, $5\times 10^3$ cm$^2$/Vs (left to right, also $T_D$ values are shown at top left corner in the figures including a value for $10^3$ cm$^2$/Vs, i.e., the highest temperature in each figure). The critical density can be calculated from $E_c=k_B T_D =\Gamma$, i.e., $n_c = (g_sg_v/4\pi)(e/\mu \hbar)$, which is independent on the effective mass for a given mobility. Thus, for given mobilities $\mu = 10^3$, $5\times 10^3$, $2 \times 10^4$, $5 \times 10^4$, $2\times 10^5$ cm$^2$/Vs, we have the critical densities $n_c=48.3$, 9.7, 2.4, 0.97, 0.24$\times 10^{10}$ cm$^{-2}$ for $g_v=2$, and  $n_c=144.9$, 29.0, 7.2, 2.9, 0.72$\times 10^{10}$ cm$^{-2}$ for $g_v=6$. For different mobility and valley degeneracy $n_c$ scales as $n_c \propto g_sg_v/\mu$.
(d) Resistivity of a GaAs system as a function of temperature for different densities (in the unit of $10^{10} cm^{-2}$), which are calculated with the parameters corresponding to GaAs with $g_v=1$ and the interface impurities between GaAs and GaAlAs. 
The vertical red dashed lines indicate the Dingle temperatures corresponding to the mobility $\mu = 10^7$, $10^6$, $10^5$ cm$^2$/Vs (left to right, also $T_D$ are shown at top left corner in the figures including values for $\mu=10^4$ and $10^3$ cm$^2$/Vs, i.e., 10K and 100K, respectively). For given mobilities $\mu = 10^7$, $10^6$, $10^5$, $10^4$, $10^3$ cm$^2$/Vs, the critical densities are given by $n_c=0.002$, 0.024, 0.242, 2.42, 24.2$\times 10^{10}$ cm$^{-2}$ for $g_v=1$, respectively.
Values of $T_D$ ($n_c$) indicate the temperature (density) thresholds above which the metallicity behavior may manifest itself as discussed in the text.
}
\end{figure}

To establish the qualitative validity of our analytical results presented above, we provide in Fig.~4 our detailed numerically calculated results for the 2D resistivity in 2D Si(100), Si(111) (using both $g_v=2$ and 6), and n-GaAs ($g_v=1$) systems directly using Eqs.~(\ref{eq1})--(\ref{eq6}) with no additional approximations. It is manifestly clear that the numerical results establish that the metallicity is the strongest (weakest) in Si(111) (n-GaAs) system exactly as our analytical considerations imply. In Fig.~4, we have shown by vertical lines various values of the mobility-dependent Dingle temperature ($T_D$) which would cut off the temperature dependence, explicitly bringing out the fact that the low-mobility samples with high disorder (and the associated high $T_D$ values) would not manifest any metallic behavior. An alternative statement is that low mobility implies high values of metallic density
in Fig.~4 applying only for $n>n_c$ 
where the metallic temperature dependence is weak.  This
means that the observation of any metallic behavior necessitates low values of $n_c$
where Fig. 4 shows strong metallic temperature dependence

We conclude this section by summarizing our finding for the materials dependence of 2D systems manifesting strong 2D MIT behavior (i.e., a strong metallic temperature dependence with large $d\rho/dT >0$ for $n \agt n_c$). We find that $T_c$ defined by $T_c = T_F(n=n_c)$ must be small enough so that $T_c < T_{BG}$ for phonon effects to be negligible at low temperatures. We also need disorder to be small enough so that $T_D < T_c$, and therefore $T_D < T_c < T_{BG}$ must be satisfied as the necessary condition for the manifestation of 2D MIT. The sufficient condition is given by $q_{TF}/2k_F > 1$ (or at least, not too small) for $n > n_c$ so that $d\rho/dT$ is not too small.  Using the known expressions for the relevant variables $T_{BG}$, $T_F$, $T_D$, $q_{TF}$, and $k_F$ we conclude that $g_v$, $m$, and $v_{ph}$ should be as large as possible [see Eq.~(\ref{nc})], disorder should be as small as possible so that $\mu$ is large [see Eq.~(\ref{muc})], and $g_v^3 m^2/\kappa^2$ should be as larger as possible [see Eq.~(\ref{ncm})], implying not only large $g_v$ and $m$, but also small $\kappa$. This immediately leads to the conclusion that high-mobility Si(111) 2D systems will manifest the strongest 2D MIT behavior (since $g_v=6$ here, and $m$ is large) whereas 2D n-GaAs will have the weakest 2D MIT behavior (since $m=0.07m_e$ is the smallest here with $g_v=1$), and high-mobility Si(100), 2D p-GaAs, and 2D SiGe systems will have strong 2D MIT behaviors since $m=0.19$ and $g_v=2$ [for Si(100)], $m=0.4$ and $g_v=1$ (for 2D p-GaAs) are consistent with strong 2D MIT behavior. It is gratifying to know that this 
material-dependence is exactly what is manifested experimentally with high-mobility Si(111) 2D systems showing \cite{kane111,hwangprb2007} the strongest 2D metallic behavior and 2D n-GaAs showing the weakest 2D metallic behavior \cite{lillyprl2003}. Of course, if the sample mobility is low so that the condition $T_D < T_c < T_{BG}$ is violated, then there would be no 2D MIT behavior at all, as happened in almost all low-mobility 2D systems prior to 1995. We also mention that if the spin degeneracy is lifted (so that $g_s=1$ instead of 2), for example, by the application of a strong parallel magnetic field, then 2D MIT behavior is suppressed according to the above considerations, and as observed experimentally.
Our presented numerical results (Fig. 4) agree with our analytical results.

\section{disorder and interaction effects on 2D polarizability} 

The mean-field RPA-Boltzmann screening theory approach to disorder-limited 2D transport used in our analyses so far ignores the effects of disorder and interaction on the screening function itself, and incorporates the temperature-induced modification (Fig.~\ref{fig1}) of the finite wave number 2D polarizability function as the key physical mechanism controlling the observed intrinsic metallic behavior.  The two dimensionless parameters controlling the metallic temperature dependence of 2D transport are $T/T_F$ and $q_{TF}/2k_F$, both of which should be large (or at least, not too small) for the manifestation of metallicity.  This immediately leads to the question of how disorder and interaction themselves modify the 2D screening function and whether the temperature dependence of the resistivity arising from the thermal suppression of $2k_F$-screening in the noninteracting 2D polarizability function of the clean system is theoretically robust beyond the zeroth order mean field RPA-Boltzmann theory used in our considerations.  This is of course an important, but also a very hard, open question whose answer can at best be approximate in any attempted theories since the fate of an interacting electron system (either in the continuum jellium model of an electron liquid as appropriate for our system or in the corresponding Mott-Hubbard-Anderson model on a lattice) in the presence of disorder is unknown as the problem is a true strong-coupling non-perturbative problem (with the notorious fermionic sign problem not amenable to large scale computer simulations).  

In this section, which should be considered a continuation of the section II of our article, we present some simple calculations going beyond the RPA theory of screening including disorder and interaction effects, and arguing that perhaps disorder and interaction, when they are not too strong, would not change the picture qualitatively, but our claims and findings in this context are rather modest and should be taken as very approximate attempts toward a long-standing unsolved problem.  There are alternative (and more ambitious) theoretical approaches \cite{zalaprb2001,punnoose,spivak}
to the problem (of including disorder and interaction in the 2D transport theory) in the context of 2D MIT phenomena which are complementary to our work (and which also happen to be much less predictive
than our theory -- the great advantage of our zeroth order theory is its simplicity enabling us to make precise quantitative predictions as described in section III of this article).  We mention that the consideration of interaction effects (beyond RPA screening) is not just of academic interest here since the physics of the 2D metallic behavior (i.e., the manifestation of a strong  temperature dependence in the 2D resistivity) is inherently a low-density phenomenon (by virtue of the necessity of $x=q_{TF}/2k_F$ being not too small and $n_c$ being low so that $T_F$ is not too high) as emphasized in the last section.  In fact, we can rewrite the dimensionless parameter `$x$' as:
\begin{equation}
 x=q_{TF}/2k_F= g^{3/2}r_s \sim n^{-0.5},
\end{equation}
where, $r_s=me^2/\kappa \sqrt{\pi n}$
is the dimensionless Wigner-Seitz radius characterizing the interaction strength in an electron liquid and $g=g_s g_v$ is the total ground state degeneracy.  We note that a large (or not too small) value of $x$, as necessary for strong 2D metallicity, implies that $r_s$ cannot be too small which then brings into question the quantitative validity of the RPA screening approach since $r_s$ should be small ($r_s<1$) for the quantitative validity of RPA.  Thus, some justification is needed in ignoring interaction effects in a low-density ($r_s>1$) electron system where $1/r_s$ basically defines the average number of electrons participating in a typical screening cloud.  The main justifications for our mean-field RPA approach (other than its simplicity and predictive power) are (1) RPA is empirically known to work well for the quantitative description of many interacting Fermi liquid properties in 3D metals which typically have $r_s>5$, \cite{mahan,fetter}
and (2) perhaps even more importantly, the 2D MIT phenomenon is primarily a ``high-temperature" phenomenon where $T/T_F$ cannot be too small for the observation of metallic behavior, and as such, interaction effects might not be too crucial.  
In particular, $T>T_D$ ($\hbar/tau$) is necessary for 2D metallicity to manifest itself, making the phenomenon essentially a high-temperature phenomenon where quantum correlation effects might be small.
Nevertheless, it is necessary to investigate both disorder and interaction effects on the 2D screening properties going beyond RPA which is what we do below.

In this section, we theoretically consider the influence of random charged impurity
disorder \cite{disorder} and exchange-correlation effects arising from
electron-electron interactions \cite{interaction} on the 2D electron
polarizability (or equivalently the screening function). 
The noninteracting static
2D screening function in a clean system was first calculated by  Stern \cite{sternprl1967} within the random phase approximation (RPA). In
the current work we consider two separate generalizations of the RPA
screening theory: Inclusion of impurity scattering effects and
inclusion of electron-electron interaction effects. In both cases, we
consider theoretical approximation schemes to go beyond the
simple RPA theory. 
The full transport theory including both disorder and interaction effects on screening, however, remains a formidable open challenge for the future well beyond the scope of the current work although we cite approximate efforts in this direction by other groups using alternative 
(and highly approximate)
theoretical techniques \cite{zalaprb2001,punnoose,spivak}

Our reason for considering (separately) both disorder and interaction
effects on electronic dielectric screening is simply the fact that
both become important in the low carrier density regime, and thus it
is important to have some approximate estimate of both corrections
to the basic RPA screening theory. In this context it is important to
emphasize that we have assumed throughout that the system remains
homogeneous even in the low carrier density regime so that the
standard ensemble averaged diagrammatic perturbation theory is
applicable even in the presence of disorder. 
This may not, however, be true in the low carrier density regime in the presence of long-range Coulomb disorder where linear screening itself may fail due to the non-perturbative formation of charged impurity-induced electron puddles leading to an inhomogeneous density landscape which we have discussed elsewhere \cite{liprb2013}.
In the current work, we assume that the system remains homogeneous throughout and consider disorder and interaction effects on the electronic polarizability function diagrammatically.

\begin{figure}[t]
	\centering
	\includegraphics[width=.9\columnwidth]{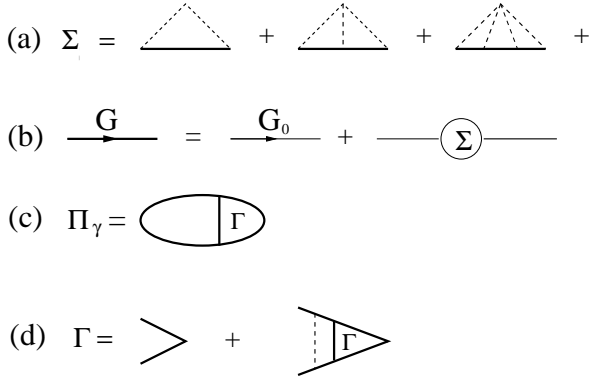}
\caption{
Set of diagrams used in this calculation: (a) the self energy corrections, (b) renormalized Green function, (c) static polarizability function ($\Pi_{\gamma}$) formed from the 
renormalized Green function, and (d) impurity ladder vertex corrections.
\label{fig5}
}
\end{figure}

\begin{figure}[t]
	\centering
	\includegraphics[width=.9\columnwidth]{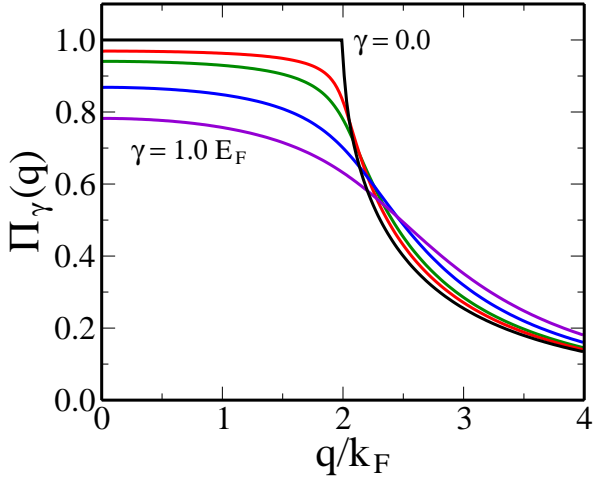}
\caption{The calculated polarizability as a function of wave vector for various scattering rate $\gamma=0$, 0.1, 0.2, 0.5, 1.0$E_F$ (from top to bottom).
}
\label{fig6}
\end{figure}

We first consider disorder effects on the static 2D polarizability.
Since Coulomb
disorder with its long-wavelength divergence must necessarily be
screened for meaningful results, the inclusion of disorder effects on
screening involves a nonlinear self-consistent theory where the
screened Coulomb disorder arising from random quenched charged
impurities in the background both determines screening and is
determined by it through the renormalized Green's function. 
Thus, to calculate the screening function in the presence of disorder we use the renormalized 
electronic Green's function due to the electron-impurity interaction.
The electronic self-energy corrections due to the impurity scattering 
are obtained in the non-crossing multiple-scattering approximation [see Fig.~\ref{fig5}(a)].
To calculate the self-energy corrections ($\Sigma$)
the actual electron-impurity Coulomb interaction should be 
used, but the real calculation with long range Coulomb interaction is
intractable, particularly because of the non-linear self-consistency requirement.
In this calculation we approximate the disorder 
to be of the short-ranged form, $v_0\delta(x-x_0)$.

The static polarizability function ($\Pi_{\gamma}$) formed from the 
renormalized Green's function is obtained from the ladder vertex 
Bethe-Salpeter integral equation which can be exactly solved for the short-range disorder model
[see Fig.~\ref{fig5}(c) and (d)]
\begin{equation}
\Pi_{\gamma}(q) = N_F \int \frac{d\omega}{2\pi i} \frac{\Pi(q,\omega)}
{1-\frac{\gamma}{2\pi}\Pi(q,\omega)},
\end{equation}
where $N_F = g_s g_v m/2\pi$ is the 2D density of states at the Fermi energy,
$\gamma = 2\pi n_i v_0^2 N_F$ with an impurity density ($n_i$)
is the disorder scattering strength
at the Fermi surface, and $\Pi(q,\omega)$ is given by
\begin{eqnarray}
\Pi(q,\omega)& =& N_F^{-1}\sum_k G(k,\omega)G(k+q,\omega) \nonumber \\
& = & \frac{2}{q^2}\frac{1}{iF(q,\omega)} \ln 
\frac{F(q,\omega)+i}{F(q,\omega)-i},
\end{eqnarray}
where $F(q,\omega)=(2/q)\sqrt{w+\mu-\Sigma(\omega)-q^2/4}$.
The chemical potential $\mu$ is calculated self-consistently so that, 
as $\gamma$ is changed, the total density $n=g_sg_v\int \frac{dk}{2\pi} f(k)$ is kept
constant, where $f(k)$ is the momentum distribution function in the presence of disorder and given by $f(k) = \sum_{\omega} {\rm Im} G^{-1}(k,\omega)$
with $G$ being the electron propagator including disorder scattering effects 
[Fig.~\ref{fig5}(a) and (b)].
We note that the disorder dependent chemical potential decreases approximately linearly with the scattering strength [i.e., $\mu(\gamma)/E_F = 1 - a \gamma/E_F$]. This behavior is quite a contrast  to the temperature dependence of the chemical potential, in which the chemical potential decreases with temperature exponentially at low temperatures [i.e., $\mu(T)/E_F = 1 - T/T_F \exp(-T_F/T)$].
In Fig.~\ref{fig6} we show the calculated static polarizability as a function of wave vector for various scattering rates. Since the chemical potential decreases linearly with disorder strength the polarizability at $q=0$ also decreases linearly with disorder strength. Note that the suppression of the  polarizability due to thermal effects is exponential, i.e., ($\propto \exp[-T_F/T]$).  
As shown in Fig.~\ref{fig6} the sharp cusp at $q=2k_F$ is significantly softened  by disorder 
very similar to the softening by
finite temperature effects shown in Fig.~\ref{fig1}. Thus, the sharp cusp in the electronic polarizability function at $q=2k_F$ is rounded by disorder effects in a way similar to thermal effects,
and therefore, depending on whether temperature or disorder is stronger, screening will be suppressed either by temperature or by disorder.  This immediately leads to our physical argument in section III for why $T_D$ might cut off the metallicity, (i.e., for $T<T_D$), the metallic temperature dependence of 2D resistivity will be suppressed by disorder since $T_D$ basically is a measure of the disorder strength.  This explains (at least partially) why $T>T_D$ is necessary for the manifestation of the 2D metallic behavior or equivalently why 2D metallicity is necessarily a high-temperature phenomenon (or equivalently, why disorder or $T_D$ must be very small in order for the 2D effective metallic phase to manifest itself) as argued in section III heuristically.

\begin{figure}[t]
	\centering
	\includegraphics[width=.9\columnwidth]{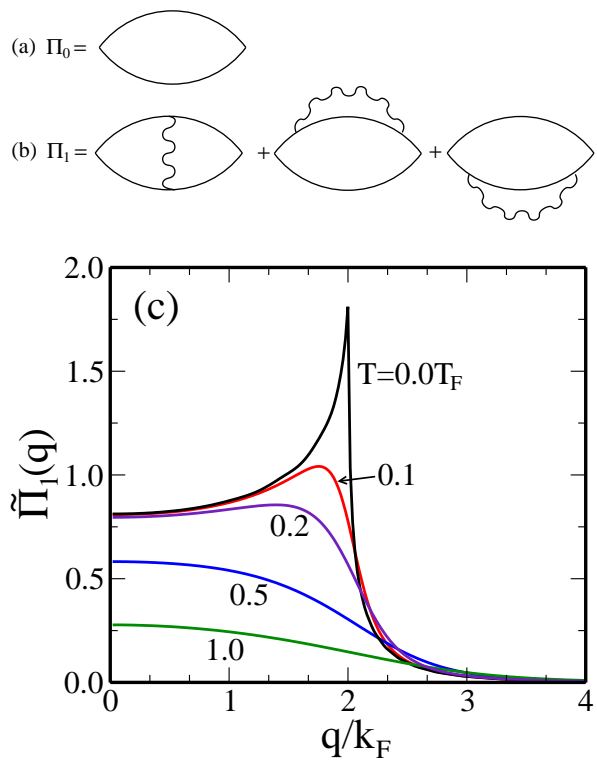}
\caption{
The diagrammatic representation of  (a) the zeroth ($\Pi_0$) and (b) the first order ($\Pi_1$) corrections to the polarizability of an interacting system. The wiggled lines in (b) indicate the bare Coulomb potential, $v(q) = 2\pi e^2/q$. 
(c) The calculated first order correction to the polarizability, $\tilde{\Pi}_1(q) = \Pi_1(q)/N_F$, as a function of wave vector for various temperatures, $T/T_F = 0.0$, 0.1, 0.2, 0.5, 1.0. 
}
\label{fig7}
\end{figure}

Now we discuss interaction effect on the 2D polarizability also using a perturbative approach. 
So far we have considered the zeroth order polarizability or the non-interacting system [i.e., bare bubble of Fig.~\ref{fig7}(a)]. In section II we discussed the temperature dependence of the zeroth order polarizability. We now obtain the finite temperature polarizability function by going to first order 
in Coulomb interaction in the
diagrammatic perturbation theory as shown in Fig.~\ref{fig7}(b). 
There are three diagrams with one Coulomb line in the first-order interaction correction to the polarizability. 
In Fig.~\ref{fig7}(c) we show the calculated total result including all three first-order diagrams. We note that although each diagram in Fig.~\ref{fig7}(b) diverges logarithmically,  the sum of the 
three first-order diagrams converges
which gives us the finite-temperature  2D static polarizability up to leading order in Coulomb interaction.
At zero temperature we have $\Pi_1(q=0) = N_F \sqrt{2}r_s/\pi$, where $N_F=\Pi_0(q=0)$ is the 2D density of states and $r_s = me^2/(\kappa \hbar^2 \sqrt{\pi n})$ is the Wigner-Seitz
parameter. In the numerical results shown in Fig.~7
we use $r_s=1.8$ which corresponds to the electron density of $n=10^{11}$ cm$^{-2}$ for the n-GaAs system. The zero temperature $\Pi_1(q,T=0)$ shows a very sharp peak at $q=2k_F$, which is a direct consequence of the 2D characteristic scattering arising at the Fermi surface.
However, as shown in Fig.~\ref{fig7} this sharp peak at $q=2k_F$ is significantly softened (suppression of the Kohn anomaly as shown in the zeroth-order polarizability in Fig.~1) by finite temperature effects. 

The interaction correction to the finite-temperature polarizability presented in Fig.~7 implies a much stronger thermal suppression of  $2k_F$-screening than in the corresponding non-interacting case shown in Fig.~1 mainly due to the strong exchange-induced enhancement of $2k_F$-screening at zero temperature.  Such a strong thermal effect would imply a very strong metallic behavior in the 2D resistivity which is not observed experimentally.  We also mention that there is not much evidence for the strong zero-temperature exchange-enhanced 2D Kohn anomaly apparent in Fig.~7(b) at T=0.  One possibility is that higher-order interaction corrections would cancel out (at least partially) the strong correction indicated by the first order effects.  The other possibility, which becomes obvious when we compare Figs.~1(a), 6, and 7(c), is that in realistic 2D systems, where both disorder (Fig.~6) and interaction (Fig.~7) are present, the two effects cancel each other out (at least partially), with interaction enhancing the thermal effect (Fig.~7) and disorder (Fig.~6) suppressing the thermal effect, leading to the original RPA screening result shown in Fig.~1 and used throughout our transport theory being a reasonable approximation.

Since the collisional level broadening (disorder) smears out the cusp at $q=2k_F$ of the zeroth order polarizability, we also expect the suppression of the sharp peak (the Kohn anomaly) in the first order correction of the screening function. Thus, the sharp peak in the electronic polarizability function at $q=2k_F$ is rounded by both disorder effects and thermal effects. 
Our general finding is that both thermal effects and impurity scattering
effects always weaken 2D screening by rounding off the singular behavior at $2k_F$ \cite{dassarmaprb1986} whereas interaction by itself could
enhance screening.

Much more work will be necessary for a complete understanding of both disorder and interaction effects on 2D transport properties, but our approximate results presented in this section indicate (at least) the possibility of a partial cancellation between the two effects, restoring some confidence in our use of RPA screening in understanding 2D transport behavior in realistic semiconductor structures where both disorder and interaction effects are undoubtedly present.

\section{discussion and conclusion}

In this work, we have investigated the necessary and sufficient conditions for different 2D semiconductor systems to manifest strong effectively metallic temperature dependence in their electrical resistivity arising entirely from intrinsic electronic mechanisms in the 
Bloch-Gr\"{u}neisen temperature regime where phonon effects are negligible. Using a physically motivated mean-field model of RPA-Boltzmann transport theory, where the carrier resistivity arises entirely from scattering off random quenched charged impurities in the environment, we have argued that the strong temperature dependence of $2k_F$-screening in 2D systems could by itself produce a metallic temperature dependence in qualitative agreement with experimental findings in many 2D systems.  We have shown that such a temperature-dependent $2k_F$-screening, and therefore the 2D metallic behavior manifested in the strong temperature-dependent resistivity, is intrinsic to 2D systems and does not happen in 3D metals or semiconductors which manifest the usual Sommerfeld thermal behavior.  Using the screening theory, we have derived a number of simple analytical results, supported by direct numerical results, putting constraints on the temperature, density, and mobility regimes where the 2D metallic behavior would be most pronounced and provided a reasonable explanation for why such a 2D metallic behavior was not observed during the numerous studies of 2D transport in Si-based 2D systems in the 1970s and 1980s.  Finally, we considered through leading-order perturbative approximations how the inclusion of disorder and interaction effects (i.e., going beyond the mean field RPA theory) could modify 2D screening properties, arguing that the two effects, while being strong individually, oppose each other qualitatively and quantitatively so that RPA may not
necessarily be a bad approximation for understanding transport properties of 2D semiconductor systems.

The motivation for the current work is easy to state.  Imagine someone has a 2D sample of some material [e.g., a Si(100) MOSFET] where the only experimental information available is the value of maximum mobility in the sample at low temperatures (or just the critical density for the system to become a strongly localized insulator at low temperature).  Can we predict the 2D metallic behavior in terms of the density and temperature regime where a strong temperature-dependent resistivity would manifest itself?  Our analyses indicate that the answer to this question is affirmative.  In particular, the necessary condition for the 2D metallicity to manifest itself is that the inequality 
$T_D < T_c < T_{BG}$ must be satisfied where all quantities are defined in Sec.~III and 
$T_c=T_F$ ($n=n_c$),  whereas the sufficient condition is that the value of the dimensionless parameter $x=q_{TF}/2k_F$ must not be small at the density 
being used for
the resistivity measurement. This latter condition implies that the experimental density $n$ ($>n_c$) for the observation of the metallicity must satisfy the inequality $n_c < n < n_M$ where $n_M$ is defined in Sec.~III.  Our analytical considerations immediately lead to the conclusion that the 2D metallicity would be the strongest in the 6-valley degenerate Si(111) 2D system, intermediate in Si(100) and p-GaAs 2D systems, and by far the weakest in 2D n-GaAs system where one would have to go down to $n \approx 10^9$ cm$^{-2}$ density for observing any appreciable metallic temperature dependence in the resistivity necessitating mobilities around $10^7$ cm$^2$/Vs.  All of these conclusions are consistent with our detailed numerical results and experimental results from many laboratories.

Before concluding, we want to discuss two salient features (one occurring at low temperatures, 
$T \ll T_F$, and the other at high temperatures, $T \sim T_F$) of our screening theory with concrete and falsifiable predictions.  First, the theory [see Eq.~(\ref{rholow})] predicts a maximum possible magnitude of the 2D metallicity as defined by the dimensionless parameter $d\tilde{\rho}/dt$ where $\tilde{\rho} = \rho (T)/\rho_0$ where $\rho_0= \rho(T=0)$ and $t=T/T_F$.  At very low temperature ($T \ll T_F$), Eq.~(\ref{rholow}) gives:
\begin{equation}
d\tilde{\rho}/dt= 2x/(1+x),
\end{equation}
where $x=q_{TF}/2k_F$. 
Assuming a very clean system so that $T_D \ll T \ll T_F \ll T_{BG}$ (and no complications from weak localization corrections ignored in our consideration), the above formula predicts that the dimensionless temperature derivative of a 2D system must vary as  $ \sim x  \propto n^{-0.5}$ at high density ($x \ll 1$) and become just `2' asymptotically at low density ($x \gg 1$).  The clear prediction is that all 2D systems would have a maximum possible value of two for the dimensionless metallicity as defined by the temperature derivative $d\tilde{\rho}/dt$!  This is a concrete prediction which should be experimentally checked for different 2D systems by carrying out transport measurements for the best possible samples (highest mobility) at the lowest possible temperatures and densities.  To the best of our knowledge, this quantitative prediction has never been experimentally tested. 
Any experimentally measured deviation from this predicted RPA metallicity would indicate the quantitative importance of interaction or disorder, providing clear directions of future theoretical work in the subject.

The second prediction is the high-temperature quantum-classical crossover predicted around $T \sim T_F$ in the theory where the transport behavior changes from `metallic-like' ($d\rho/dT>0$) to `insulating-like' ($d\rho/dT<0$) at some density-dependent finite temperature $T^*$ with the asymptotic analytical properties given respectively in Eqs.~(\ref{rholow}) and (\ref{rhohigh}) for $T \ll T_F$ and $T \gg T_F$.  Provided that the condition $T_{BG}  \gg T_F$ is satisfied at the particular density,  this temperature-dependent crossover (at a constant carrier density) is generically present in all 2D MIT experiments where the measured resistivity $\rho(T,n)$ invariably changes its sign at some high crossover temperature 
$T^*(n)$ with $T^*$ being defined by $d\rho/dT= 0$ at $T=T^*$ at a fixed density $n$.

The predictions of Eqs.~(\ref{rholow}) and (\ref{rhohigh}) respectively are that $\rho(T) - \rho_0 \sim T$ deep in the quantum regime ($ T \ll T_F$) and the conductivity $\sigma=1/\rho$ goes as $\sim T$ in the classical regime $T>T_F$.  The high temperature `classical' linear in temperature behavior of the 2D conductivity (for $T>T^*$) has indeed been reported in several experiments \cite{last},
but the topic of high-temperature (around $T \sim T^*$) transport in 2D systems has not attracted the attention it deserves perhaps because of the uncritical focus of the community on whether the density-driven transition around $n=n_c$ at low temperature is a crossover or a true quantum phase transition.  The interesting non-monotonic temperature dependence of $\rho(T)$ around $T \sim T^*(n)$ rather clearly establishes that the 2D metallic phase is an effective high-temperature metal rather than a new quantum metal, as such, much more experimental work should be devoted to the behavior of transport around $T \sim T^*$.

\begin{figure}[t]
	\centering
	\includegraphics[width=.8\columnwidth]{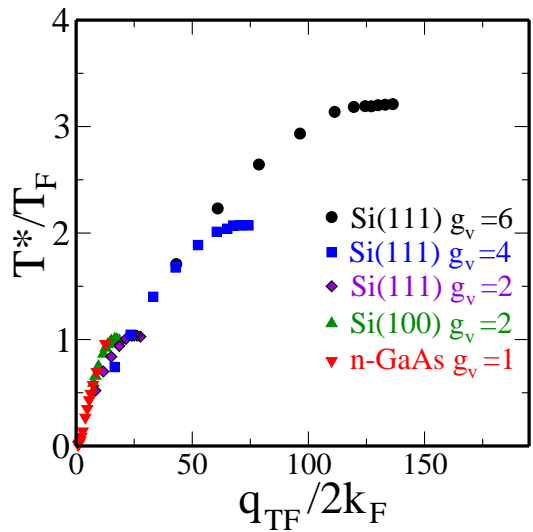}
\caption{
Numerically calculated $T^*/T_F$ as a function of the dimensionless screening parameter $q_{TF}/2k_F$ for the transport results presented in Fig.~4 which are calculated with realistic quasi-2D Coulomb form factor effect.  The figure shows the results for five distinct systems for n-GaAs with $g_v=1$, n-Si(100) with $g_v=2$, n-Si(111) with $g_v=2$, 4, and 6.
We note that for realistic parameters and carrier densities the calculated $T^*$ appears always to fall within $0.3T_F$ to $3T_F$ for all existing experimental 2D systems.
}
\label{fig8}
\end{figure}

Our numerical results presented in Fig.~4 show that the screening theory consistently gives $T^* \sim T_F$ whereas experimentally the temperature-induced crossover seems to occur over a range of temperature $T^* \sim  T_F/2 - 2T_F$, depending on the 2D material and the carrier density involved.  It is interesting that $T^*$ always falls within a factor of 2 of our predicted quantum-classical crossover temperature $\sim T_F$, but more systematic experimental work is necessary in pinning down the behavior of $T^*$ as a function of the dimensionless variable $x=q_{TF}/2k_F$ in different materials system, 
subtracting out possible phonon effects which may not be negligible at higher temperatures. \cite{mills}
Understanding the detailed quantitative behavior of $T^*$ in different systems is particularly important in view of a competing renormalization group based two-parameter 
scaling theory \cite{punnoose,punnoose2}
for the metallic phase which claims, in contrast to our contention, that 2D MIT is indeed an interaction-driven quantum phase transition with the crossover regime  $T^*<T_F$ playing a crucial role in the quantum coherent scaling properties of the system.  There has even been a claim of an experimental verification \cite{anissimova}
of the scaling theory where the resistivity data around the crossover temperature $T^*$ plays a key role in determining the quantum critical properties in spite of $T^* \sim T_F$ (i.e., an essentially classical regime) in the experiment!  To contrast our screening theory with the scaling theory, it is crucial to understand the detailed behavior of $T^*$ in different materials.  We note, however, that the scaling theory, being a theory involving an expansion in $1/g_v$, should be essentially exact for the 6-valley 
Si(111) 2D system where the experimentally observed $T^* \sim 1.5T_F$, in agreement with our theory, is in apparent disagreement with the scaling theory where the stringent requirement of $T^*<\hbar/\tau$ puts $T^*< 100$ mK in the Si(111) system in sharp contrast with 
experiments\cite{huprl2015}.  
The key difference between our theory and the scaling theory is that ours is a ballistic ``high-temperature" theory where $T>T_D$ is necessary for metallicity (and $T^* \sim T_F$ necessary for the quantum-classical crossover induced maxima in the resistivity) whereas the scaling theory is strictly a theory in the diffusive ``low-temperature" ($T<T_D$) regime so that $T^*<T_D<T_F$ is necessary. In the experimental high-mobility 6-valley Si (111) 2D system \cite{huprl2015,kane111}, $T_D \sim 100$ mK whereas $T^* \sim T_F \sim 2$ K which appears inconsistent with the scaling theory (but perfectly consistent with our screening theory),
but much more work would be necessary to settle this important question.
The strong metallic behavior of the 2D p-GaAs system is also inexplicable from the scaling theory perspective since it has $g_v=1$ whereas in our screening theory, the important parameter is $g_v m$ which is about the same in two-valley Si(100) system and one-valley p-GaAs system, indicating similar metallic behaviors in these two 2D materials in the screening theory.
We also point out that the 2D metallic phase in our screening theory is an effective `high-temperature' metal in the ballistic regime by virtue of the strong screening induced metallicity in the $T_{BG}> T_F > T >T_D$ regime with weak localization induced logarithmic insulating behavior necessarily showing up in the `low-temperature' diffusive regime ($T \ll T_D$) although such weak-localization effects may not always be easy to see experimentally due to electron heating effect necessarily present in semiconductor carriers at low temperatures.  We mention that the very low-temperature insulating upturn in the 2D resistivity has been experimentally seen deep in the diffusive regime casting doubts on the scaling theory claim of a generic quantum phase transition in 2D interacting disordered systems. 
\cite{three,weaklocal}

The calculation of $T^* (n)$, the characteristic `high' ($ \sim T_F$) temperature where the resistivity derivative changes its sign from being metallic-like ($d\rho/dT>0$) to being insulating-like ($d\rho/dT<0$), is straightforward, if somewhat tedious, in our screening theory as we need to solve the integro-differential equation defined by the condition:
\begin{equation}
d\sigma(n,T)/dT= 0,
\end{equation}
using the integrals given in Eqs.~(1)--(6) defining the temperature and density dependent conductivity, $\sigma=1/\rho$.  First, we note that it follows directly from Eqs.~(1)--(6) and Eq.~(29), neglecting all nonessential complications arising from quasi-2D wavefunction and impurity distribution effects, that $T^*(n)$ has an approximate scaling form 
for a given material
in the strict 2D limit going as:
\begin{equation}
T^*/T_F = F (x),
\end{equation}
where $F(x=q_{TF}/2k_F)$ is a scaling function of the dimensionless screening parameter $x$ and $T_F$ sets the electronic energy scale in the problem.  
Of course in real systems, the quasi-2D carrier wavefunctions, the actual impurity distribution in the sample, and possible presence of additional scattering mechanisms (not included in our Coulomb disorder model) will lead to deviations from the perfect scaling form given in Eq.~(30), but an approximate scaling form is still expected.  In Fig.~8 we present our directly numerically calculated $T^*/T_F$ as a function of the dimensionless screening parameter $x$ for the transport results presented in Fig.~4 including realistic quasi-2D Coulomb form factor effects.  The calculations carried out for five distinct systems in Fig.~8 (n-GaAs with $g_v=1$, n-Si(100) with $g_v=2$, n-Si(111) with $g_v=2$, 4, 6) using realistic density values clearly establish the rather impressive scaling of $T^*/T_F$ as a function of $q_{TF}/2k_F$ ($\sim g_vm/\kappa n^{1/2}$).  Although these systems encompass very different effective masses [varying from 0.07 for GaAs to 0.3 for Si(111)] and very different valley degeneracies, the characteristic temperature $T^*$, when scaled with respect to the Fermi temperature ($T_F \sim  n/g_v m$) of the appropriate system, shows a universal scaling form with respect to $q_{TF}/2k_F$. 
This amazing scaling property of $T^*$
in the screening theory,
which has not before been pointed out in the literature, can be directly experimentally tested by measuring $T^*(n)$ in various 2D systems as a function of temperature.  We mention that our finding that $T^*/T_F$ essentially falls between $0.5T_F$ and $2T_F$ (as can be seen in Fig.~8) for most 2D experimental systems,  and that $T^*/T_F$ tends to be larger in the Si(111) $g_v=6$ system (around $1.5-2T_F$)\cite{huprl2015}
compared with n-GaAs system ($T^*\sim 0.5T_F$) \cite{lilly} with the Si(100) 2D system being intermediate is consistent with experimental results.  In addition, our finding of $T^*/T_F$ increasing with $q_{TF}/2k_F$, which implies that $T^*/T_F$ should increase (decrease) with deceasing (increasing) carrier density in the same sample is also consistent with experiments.  It is curious that both our screening theory and the interaction-based RG theory \cite{punnoose,punnoose2}
 predict $T^*$ to be a scaling function, albeit for qualitatively different reasons with ours being a `high-temperature' ballistic theory for $T^*$ ($\sim T_F$) and the RG theory being a `low-temperature' ($T^* \ll T_F$) diffusive theory.


Finally, we comment on the possibility of  screening-induced metallic temperature dependence in 2D graphene or other related chiral systems where also charged impurity scattering is often the dominant resistive scattering mechanism \cite{graphene}.
It turns out that the chiral nature of graphene completely suppresses 2D backscattering (i.e. essentially no $2k_F$-scattering in graphene), and therefore, one does not expect \cite{falko}
any screening-induced metallic temperature dependence in graphene (or 3D topological insulator surface layer) resistivity, and in fact, very high-quality high-mobility graphene manifests weakly insulating temperature dependence at low carrier densities and low temperatures,  which has been experimentally verified \cite{graphene2}.

\section*{Acknowledgements}

This work is supported by LPS-CMTC and Basic Science Research Program through the NRF of Korea  (2009-0083540).

\end{document}